%% file: main.tex
\documentclass[sigconf]{acmart}

\AtBeginDocument{%
  \providecommand\BibTeX{{%
    \normalfont B\kern-0.5em{\scshape i\kern-0.25em b}\kern-0.8em\TeX}}}

\setcopyright{acmcopyright}
\copyrightyear{2022}
\acmYear{2022}
\acmConference[WWW]{Online, hosted by Lyon, France}{April 25-29, 2022}{The Web Conference 2022}

\usepackage{subcaption}

\newcommand\eat[1]{}
\begin{document}

\title{cGraph: Graph Based Extensible Predictive Domain Threat Intelligence Platform}

\author{Wathsara Daluwatta\textsuperscript{†},  
  Ravindu De Silva\textsuperscript{†}, 
  Sanduni Kariyawasam\textsuperscript{†},  
  Mohamed Nabeel\textsuperscript{‡}, 
  Charith Elvitigala\textsuperscript{†},
  Kasun De Zoysa\textsuperscript{†}, 
  Chamath Keppitiyagama\textsuperscript{†}} 
\affiliation{ 
  \institution{\textsuperscript{†}University of Colombo School of Computing} 
  \institution{\textsuperscript{‡}Qatar Computing Research Institute}
  \country{Sri Lanka, Qatar}
}










\renewcommand{\shortauthors}{Daluwatta et al.}


\begin{abstract}
\input{abstract}

\end{abstract}


\maketitle

\section{Introduction}
\noindent
\input{introduction}

\section{Heterogeneous Network Graph}\label{sec:kg}


\input{network}

\section{System Architecture}\label{sec:archi}
\input{system}

\section{Current Threat intelligent Solutions vs cGraph}

\input{solutions}

\section{Demonstration}
\input{demo}

\section{Conclusions}
\input{conclusion}



\end{document}

%% file: abstract.tex
 Ability to effectively investigate indicators of compromise and associated network resources involved in cyber attacks is paramount not only to identify affected network resources but also to detect related malicious resources. Today, most of the cyber threat intelligence platforms are reactive in that they can identify attack resources only after the attack is carried out. Further, these systems have limited functionality to investigate associated network resources. In this work, we propose an extensible predictive cyber threat intelligence platform called cGraph that addresses the above limitations. cGraph is built as a graph-first system where investigators can explore network resources utilizing a graph based API. Further, cGraph provides real-time predictive capabilities based on state-of-the-art inference algorithms to predict malicious domains from network graphs with a few known malicious and benign seeds. To the best of our knowledge, cGraph is the only threat intelligence platform to do so. cGraph is extensible in that additional network resources can be added to the system transparently. 
 

%% file: introduction.tex
Internet domains are the launch pad for many cyber attacks we observe nowadays. One effective way to reduce the damage caused by such attacks is to identify the domains involved early and take actions such as blocking, taking down or sinkholing. Further, these malicious domains have underlying associations with other Internet resources that help to uncover malicious infrastructures. We observe that existing threat intelligence systems 
,such as Cisco Umbrella Investigate\eat{\cite{cisco}} and Anomali\eat{\cite{anomali}}, fail to take advantage of such associations to detect stealthy malicious domains nor do they capture such associations in a user-friendly graphical form that allows users to easily explore related Internet resources. All these solutions are reactive in nature and by the time security operation teams observe them the damage is already done. A predictive solution would immensely assist in mitigating such therats.  Motivated by these gaps in the industry, we  build a proactive threat intelligence platform that can uncover malicious domains early and visualize associated Internet resources in a graph format for easy analysis.

Building an investigation platform in this domain involves a unique set of challenges. First, the amount of data one needs to process on a daily basis is huge. On average, we observe around 850 million network records per day. Second, the network data is highly dynamic. For example, domains change their hosting IPs periodically, the ownership of IPs change over time, new domains are created and existing domains cease to exist. Third, intelligence sources such as Phishtank, VirusTotal and GSB, contain limited information about the status of domain and they are slow to detect malicious domains~\cite{graphinf:2020}. 
This is due to the fact that most of such systems rely on either website content or user accessing of these websites to ascertain whether they are malicious or not. Providing timely and up-to-date intelligence is thus a challenge on its own. Fourth, network data, though mostly structured, is available in different data formats and these formats over time. Fifth, collecting additional network data and intelligence seeds is often rate limited and encounter frequent network failures. Sixth,  storage and retrieval of huge amounts of graph data, especially in community databases, are not demonstrated to work reliably in practice.

In this work, we build a system called cGraph to address the above mentioned limitations. We design to implement a scalable big data processing pipeline that can ingest billions of disparate reports daily. In order to track the changes over time, we build succinct snapshots of the whole Internet on a daily basis. 
cGraph provides predictive intelligence by incorporating state of the art graph inference based malicious domain prediction techniques~\cite{graphinf:2020}. In order to make the system oblivious to data format changes, cGraph provides one level of indirection to ingest data as common schema into the system. This allows one to incorporate data format changes without requiring to change the data processing pipeline allowing to extend with additional data sources transparently. Further, unlike prior work~\cite{graphinf:2020} which utilizes static graphs to infer malicious domains, cGraph is able to construct a graph around the domain in question on the fly and generate real-time intelligence. This gives cGraph to predict the maliciousness of any domain provided there are seed domains in the neighborhood. 

\textbf{Demonstration Scenarios}: cGraph assists in domain threat investigation as well as identification of associated malicious domains~\footnote{Demonstrations of a combosquatting domain and a benign domain investigation are available at https://youtu.be/yu47CZkp\_eY and https://youtu.be/waxczBk7vA8 respectively.}. We show three use cases: (1) searching for impersonating domains and discovering related malicious domains, (2) real-time reputation score for a domain via a browser based plugin, and (3) investigation of the evaluation of network resources over time around a domain under consideration. 


%% file: network.tex
\subsection{Data Sources}
The system ingests four types of data/intelligence sources to build the graph: (1) Farsight Security PDNS~\cite{farsight} - Passive DNS data observed from all around the world which covers the majority of the DNS resolutions in the Internet., (2) VirusTotal (VT)~\cite{virustotal}  - Reputation service that provides aggregated intelligence on any URL, which is the most popular aggregated threat intelligence platform that provides threat reports using more than 80 antivirus engines, blacklists, and sandbox systems., (3) Maxmind \cite{maxmind} - A comprehensive database consisting of geolocations and ASN information of IP addresses (4) Alexa top 1 million domains~\cite{alexa} - Alexa is a widely used resource for search engine optimization and capturing the status of domains. 

In summary, we use Farsight passive DNS records to generate comprehensive knowledge graphs and use Alexa Top 1 million domains, VirusTotal, and Maxmind information to inject the graph with benign and malicious seed domains for running belief propagation and predicting new malicious domains~\cite{graphinf:2020}.

\subsection{Graph Modeling}
To create the knowledge graph (KG), we employ the DNS record types A, NS, MX, SOA, and CNAME, which include the most important information on domains. Specifically, Type A records contain domain to IPv4 address resolutions, Type NS records contain domain to nameserver resolutions, Type MX records contain domain to mail server resolutions, and CNAME records provide domain to alias domains resolutions. All records are created by having a domain name as their pivot. We create a heterogeneous KG utilizing the above DNS records with different edge types representing different record types and nodes representing domains or IP addresses. Our intuition of building such a KG is that attackers are likely to host their domains in the same infrastructure where some of their blacklisted domains are hosted. This allows one to utilize label propagation algorithms to predict malicious domains from a small seed. 



We build a heterogeneous KG based on the following different types of DNS records available in PDNS. 

\begin{itemize}
    \item Vertices: Apex, FQDNs, IPs, Name Servers, Mail Servers, CNAME, SOA
    \item Edges: Apex-IP, Apex-FQDNs, FQDNs-IP, Apex-NS, FQDNs-NS, Apex-MX, FQDNs-MX, Apex-CNAME, FQDNs-CNAME, Apex-SOA, FQDNs-SOA
\end{itemize}


It is important to store the graph optimized for lookup as cGraph should be able to retrieve the KG for inference as well as replaying of network infrastructure over time quickly. Thus, the graph information is stored along with timestamps at the granularity of a day. 

%% file: system.tex
\subsection{Overall System}


The system consists of four main layers: data processing, database and caching layer, middleware and API layer, and consumer applications. The data processing layer is the component where all the data is cleaned and processed at the first phase. The infrastructure layer orchestrates the resources to insert/update the cleaned and reshaped data into the database cluster. Caching layer handles the data caching and the middleware layer handles the real-time inference process and intermediate data processing tasks.  API layer connects the backend to the consumer products, which are web application, browser extension, and public developer API~\footnote{More information is available at \url{https://qcri.github.io/cgraph/}.}.

\begin{figure}[h]
\centering
\includegraphics[width=0.6\linewidth]{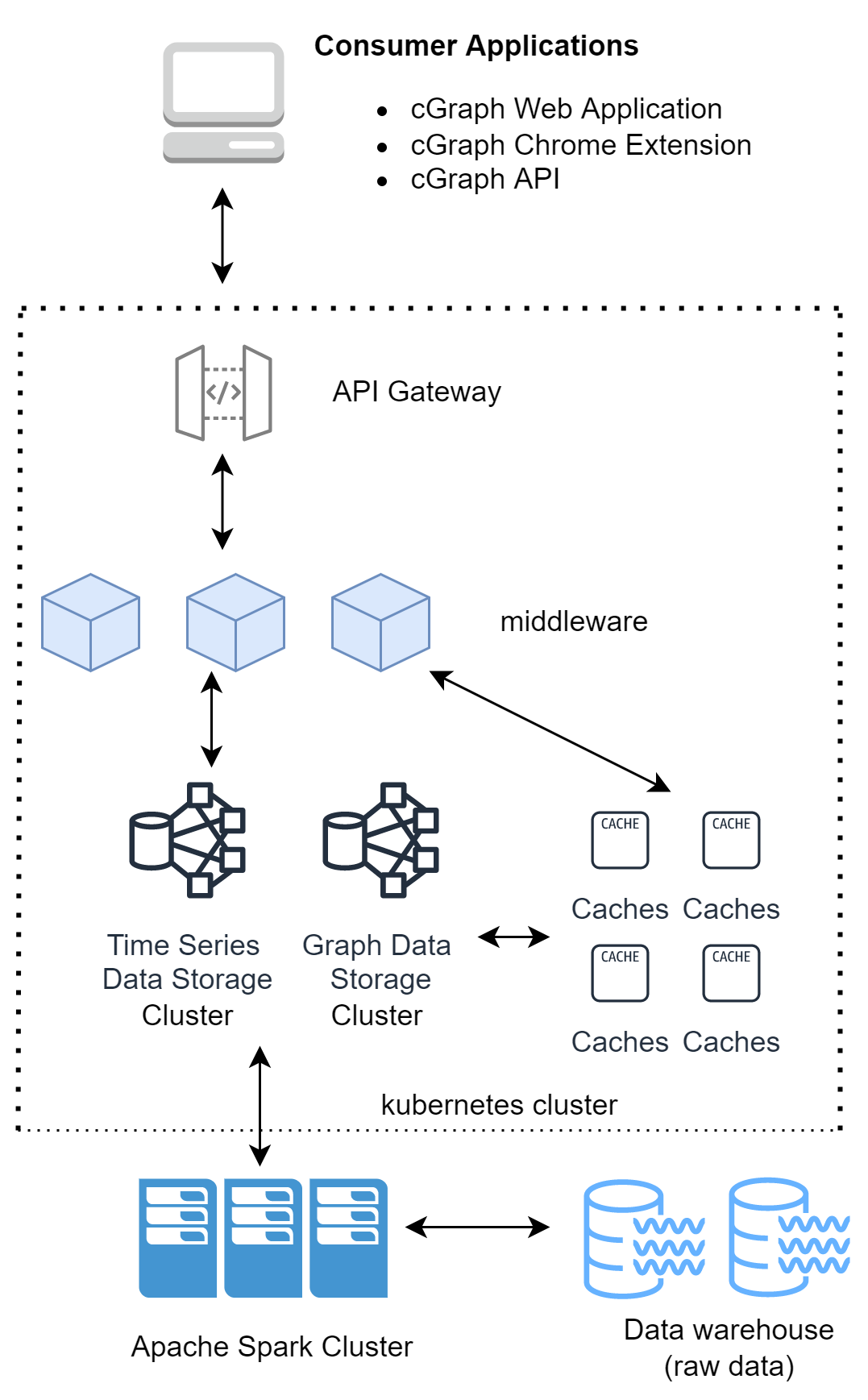}
\caption{Overall system from infrastructure point of view }
\label{fig:arch}
\end{figure}

All resources have been provisioned on top of VMware vSphere private cloud. Kubernetes cluster has been installed on top of the provisioned VMs and an abstracted computation resources layer has been created in order to manage all computation resources together.

\subsection{Graph Inference}

\textbf{Inference}. Belief propagation  (BP) is an iterative algorithm that passes the belief iteratively. In each iteration t, belief propagation algorithm updates the belief value on the node and passes the beliefs (messages) to its neighbors based on the belief values it received from the previous iteration t-1. As mentioned, this interactive process continues until each message value passed by a node to a neighbor node converges to a stable value. The final beliefs are extracted after convergence. 

\textbf{Execution}. The main idea is to construct a KG on the fly and inject ground truth labels to the KG from credible sources where VT, Alexa, and Maxmind and then run the BP algorithm on top of the knowledge graph by using the labeled values as initial beliefs. It propagates a probabilistic values for each unlabeled domain indicating the maliciousness. For bravity, we explain the execution for a selected example graph. The ground truth extraction process is conducted in two stages; benign ground truth extraction and malicious ground truth extraction. 

For the benign ground truth extraction, Alexa Rank data is used. Alexa Rank data feed for the latest seven consecutive days is taken and the domains with Alexa Rank $\leq10000$ is taken for each day. From this, the unique set of domains 
is taken as the benign ground truth. Similar to prior research~\cite{compromised}, we observe that  domains with Alexa Rank $\leq10000$ which appear consistently throughout a week are likely to be benign as malicious domains do not live longer than a day or two in Alexa 10K.

For the malicious ground truth, the data from the daily feed of VirusTotal is used. VirusTotal reports for latest seven consecutive days are taken and and the reports with a VT positive count of at least 3 is extracted as malicious ground truth as VT reports with 2 or less positive counts are known to be a less reliable indicator of maliciousness~\cite{graphinf:2020}. We manually verify random samples of benign and malicious ground truth and ascertain that they are indeed correctly labeled. 


 After seeding the KG with the ground truth domains, we run a multi-threaded version of the BP algorithm which executes within a few seconds. Our experiments show that our inference based classification achieves a high precision of 92.17\% and high recall of 99.41\%. We observe that a domain's maliciousness score does not affect significantly by considering only its two hop neighbors compared to the complete graph. Hence, for the real-time inference on an unknown domain, we build the two-hop neighborhood KG and run the inference algorithm to further improve the computational performance. 

%% file: solutions.tex
We thoroughly evaluate the key threat intelligence solutions in the market against cGraph, our threat intelligence platform: AT\&T Cybersecurity, VirusTotal Graph,  Cisco Umbrella Investigate, and Anomali Threat Intelligence Platform. Below, we provide a summary of key differentiation in our system: (1) Dynamic graph-based visualization and search of Internet artifacts for threat intelligence (all the above solutions lack an intuitive graph based navigation), (2) The ability to integrate various data sources and expand the network coverage and the ability to integrate any intelligence sources seamlessly, (3) Predictive domain threat intelligence based on an efficient graph inference algorithm (all above solutions provide only reactive intelligence), and (4) The ability to traverse the network infrastructure based on time and observe the changes in the physical layout as well as the maliciousness of the domains in the network over time, which greatly assist security operation teams in identifying root causes of attacks and the extent of the damage.

%% file: demo.tex
In this section, we show three common use cases of our system.

\textbf{Searching for impersonating domains and related other malicious domains}:
Attackers increasingly impersonate popular brands such as Apple, Paypal and Microsoft. 
cGraph allows to search for any popular brands and easily identify likely combosquatting domains.
  We demonstrate the use case of hunting for combosuqatting domains of paypal.com. 

\begin{figure}[h]
\centering
\includegraphics[width=0.9\linewidth]{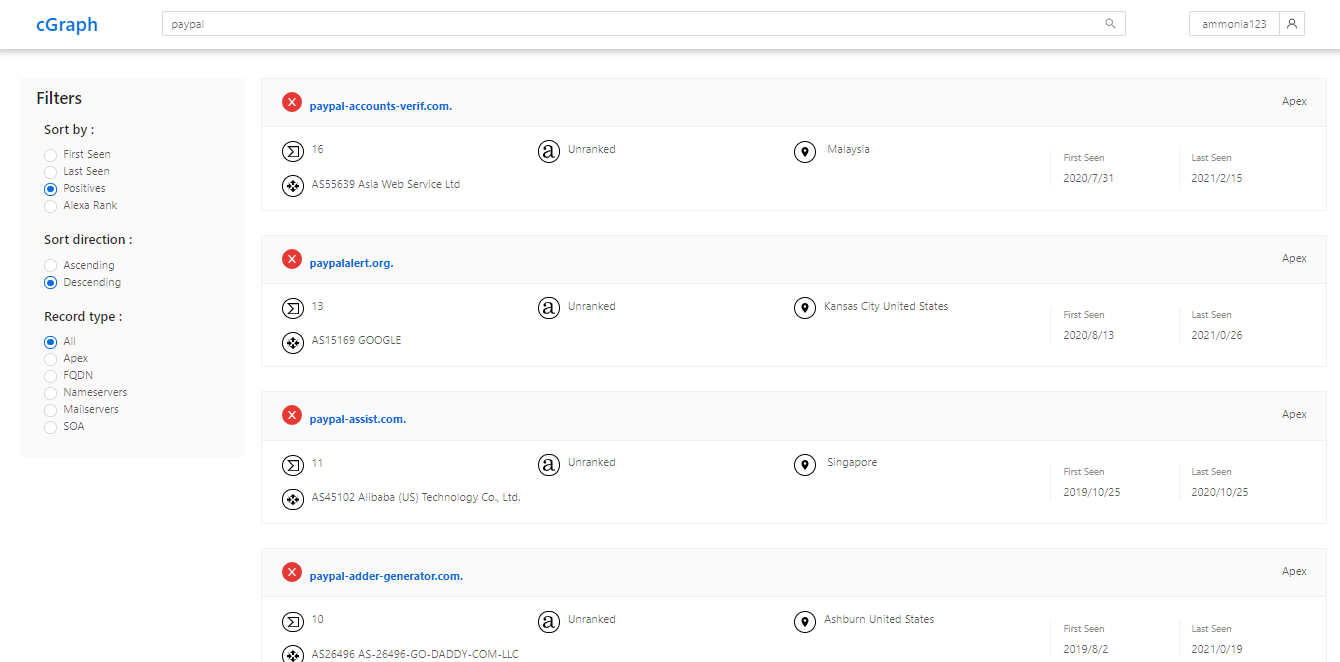}
\caption{Search results for the keyword paypal}
\label{fig:cp1}
\end{figure}

\begin{figure}[h]
\centering
\includegraphics[width=0.9\linewidth]{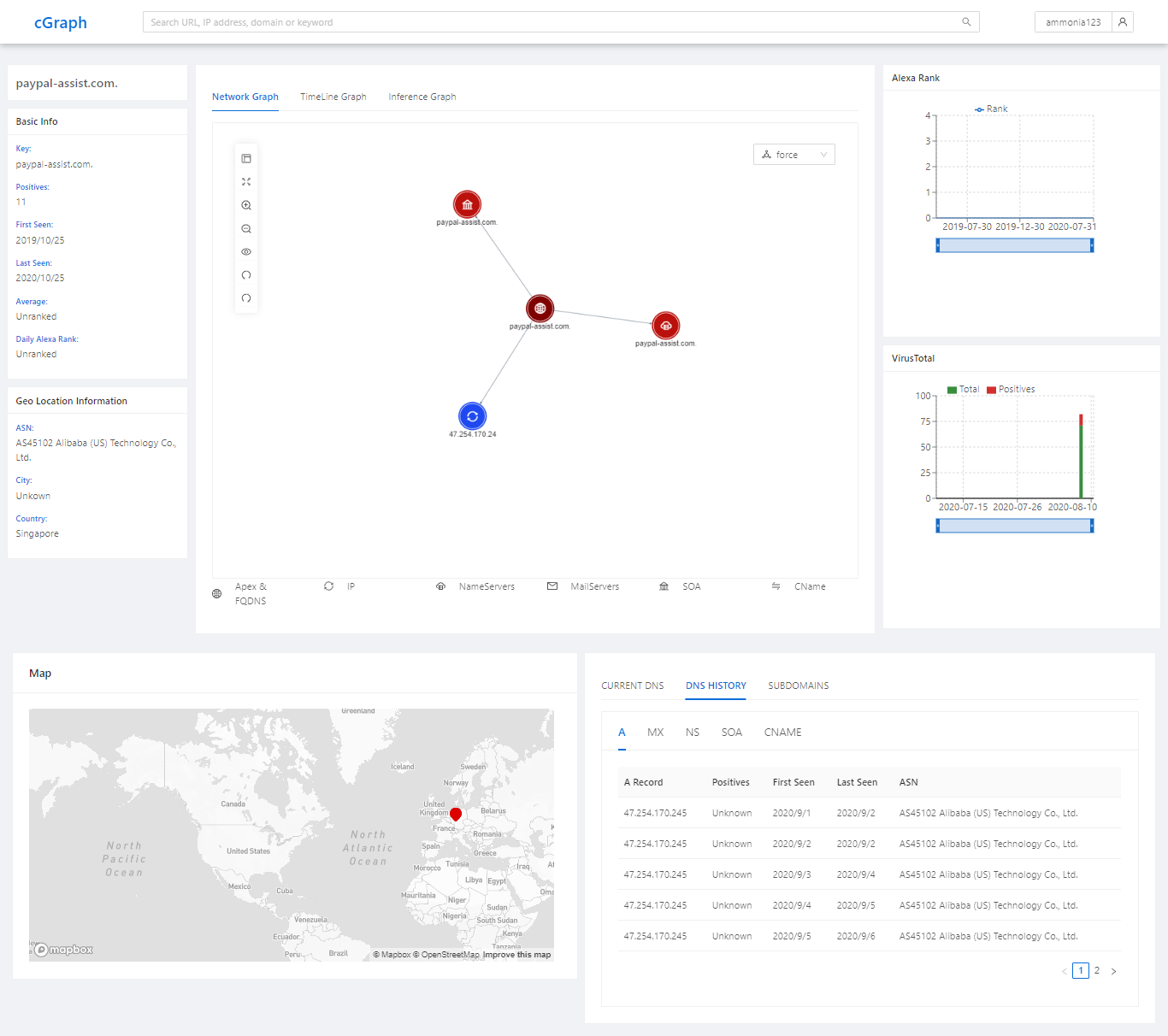}
\caption{Dashboard for the domain paypal-assit.com}
\label{fig:cp2}
\end{figure}

Figure~\ref{fig:cp1} shows the search results for the keyword paypal ranked based on VT positive counts. An investigator may pick any of these domains to further investigate. Assume that they select paypal-assist.com, which takes them to the graph based interface shown in Figure~\ref{fig:cp2}. This view includes the domain summary information, network graph, Alexa ranking and VT statistics over time, IP geolocation information and historical hosting information.

\begin{figure}[h]
\centering
\includegraphics[width=0.9\linewidth]{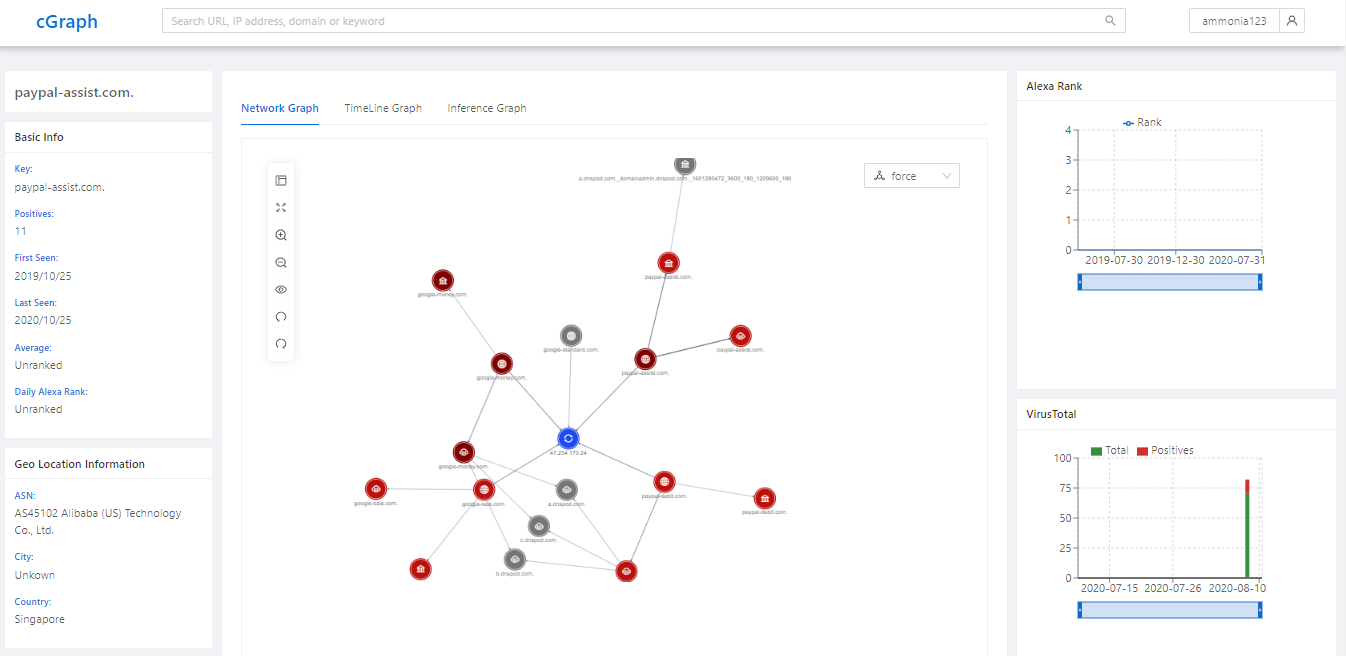}
\caption{Expansion of the IP address 47.254.170.24}
\label{fig:cp3}
\vspace{-4mm}
\end{figure}

The ability expand the network graph based on the investigator's findings is quite useful in practice. cGraph not only allows to expand any network node when it has children. but also provides real-time predict intelligence on the unknown domains in the network.
Figure~\ref{fig:cp3} shows the expansion of the IP address 47.254.170.24 that adds 4 additional domains to the graph which are hosted on this IP address. 

\begin{figure}[h]
\centering
\includegraphics[width=0.9\linewidth]{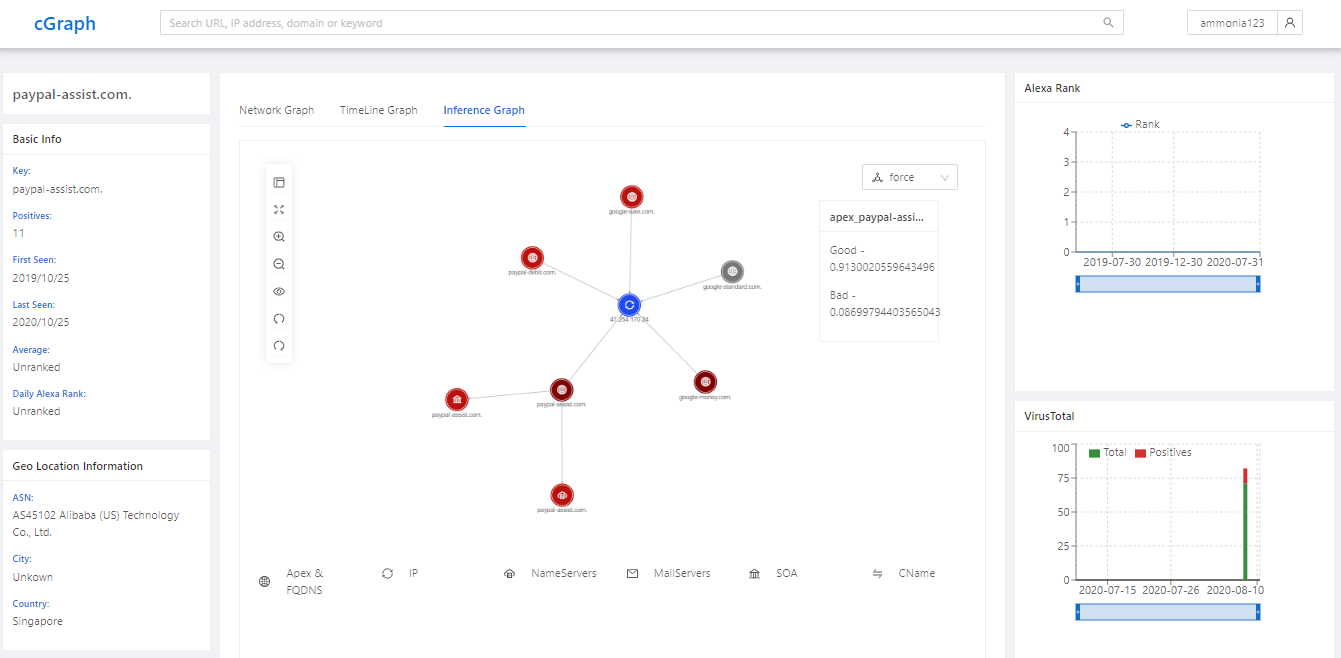}
\caption{Malicious domains identified by inference}
\label{fig:cp4}
\end{figure}

The inference graph view shows the maliciousness score of each domain in the network graph 
calculated on the fly using the BP algorithm based on known seed domains, paypal-assist.com in this case. Figure~\ref{fig:cp4} shows that cGraph inference identifies the domains google-standard.com, google-money.com, paypal-debit.com and google-sale.com to be highly malicious. 

\textbf{Evaluation of maliciousness of any domain via the browser plugin}:
In this use case, we show how cGraph API can be consumed to obtain reputation scores for websites on the fly. While blacklists such as GSB that powers the Chrome browser identify many malicious websites, they are either slow to react or unable to make a judgement if the website content is cloaked from the blacklisting service. cGraph's reputation engine takes a different approach and infers the maliciousness of domains based on the maliciousness of the neighboring network resources such as sibling domains, name servers, CNAME domains, IP addresses and certificates. As mentioned earlier, the reputation score is computed dynamically based on the status of the network infrastructure at the present moment.

\begin{figure}[h]
\centering
\includegraphics[width=0.3\linewidth]{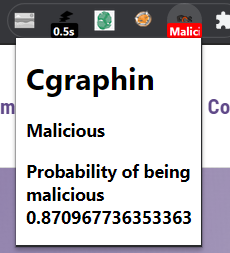}
\caption{Browser plugin: real time reputation score}
\label{fig:ce}
\vspace{-4mm}
\end{figure}

Figure \ref{fig:ce} shows the cGraph inference API in action when a user types the website soaponline.org. The maliciousness score indicates how likely the domain is malicious or benign. In this example, the maliciousness score of 0.87 indicates that soaponline.org is highly likely to be malicious.

\begin{figure}[!ht]
\centering
\includegraphics[width=0.9\linewidth]{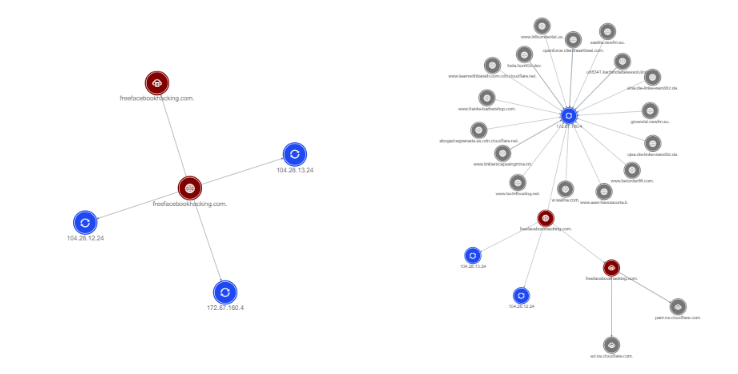}
\caption{Network timeline for day 1 (left) and day 7 (right)}
\label{fig:day1day2}
\vspace{-4mm}
\end{figure}

\textbf{Investigating the network over time}:
Network resources associated with domains change over time due to both malicious as well as benign reasons. For example, cloud IP rotations, CDN hostings and IP load balancing change IPs associated with domains frequently. Attackers also move from one IP to another as a technique to either resist take down or evade detection. cGraph timeline feature allows to investigate the network graph over a time period. Coupled with the real-time inference feature, one can observe how the reputation of any of the domains change over time. Figure~\ref{fig:day1day2} shows the network graph on day 1 and the evolved network graph on day 7 respectively. This allows the investigator to learn the fluxing behavior 
of domains and take necessary actions.


%% file: conclusion.tex
We propose, cGraph, which takes domain threat investigation a step forward. cGraph can efficiently store and process large amount of disparate network data, build network graph for any given time period and perform real-time graph inference to predict domains as malicious or benign, compared to reactive threat intelligence platforms available. We demonstrate three frequent use cases of cGraph: domain search, real-time reputation computation and network graph evolution along with the maliciousness scores of domains in the network over time.
